\documentstyle[prc,aps,epsfig,multicol]{revtex}
\tightenlines

\begin{document}
\title{Charm production from proton-proton collisions}
\bigskip
\author{Wei Liu, Che Ming Ko, and Su Houng Lee\footnote{Permanent address:
Department of Physics and Institute of Physics and Applied Physics,
Yonsei University, Seoul 120-749, Korea}}
\address{$^1$Cyclotron Institute and Physics Department, Texas A\&M University,
College Station, Texas 77843-3366, USA}
\maketitle

\begin{abstract}
We evaluate the cross sections for charmed hadron production from
proton-proton reactions $pp\to\bar D^0p\Lambda_c^+$ and 
$pp\to\bar D^{*0}p\Lambda_c^+$ using a hadronic Lagrangian.
With empirical coupling constants and cutoff parameters 
in the form factors, sum of their cross sections at center-of-mass
energy of 11.5 GeV is about 1 $\mu$b and is comparable to 
measured inclusive cross section for charmed hadron production
from proton-proton reactions. The cross section decreases to
about 1 nb at 40 MeV above threshold.

\end{abstract}

\begin{multicols}{2}

\section{introduction}

For reactions involving hadrons that consist of charm quarks, a
hadronic model with interaction Lagrangian based on the SU(4)
flavor symmetry was first introduced in Ref.\cite{matinyan}. With
empirical coupling constants and introducing form factors at
interaction vertices, this model gives a $J/\psi$ absorption cross
section by pion or rho meson \cite{haglin,lin1,oh} that is
comparable to that needed in the comover model for understanding
the observed suppression of $J/\psi$ production in relativistic
heavy ion collisions \cite{cassing1,capella}. Extending the
Lagrangian to include the interactions between charmed hadrons and baryons, 
the model has also been used to study $J/\psi$ absorption
by nucleons \cite{liu1} and charm photoproduction on nucleons
\cite{liu2}. In both cases, the theoretical cross sections are
comparable to those known empirically. The model has further been
used to calculate the cross section for charm production from
$\pi$-N interactions \cite{liu3}, which is relevant to charmed
meson production in relativistic heavy ion collisions
\cite{cassing2}, and the cross sections for charmed meson
scattering by hadrons \cite{lin2,di}. In the present paper, the
same hadronic Lagrangian is used to evaluate charmed hadron
production from proton-proton collisions. Motivated by future
experiments at proposed accelerator facility at the German Heavy
Ion Research Center \cite{gsi}, there are already studies on these reactions
based on the meson-exchange model \cite{gasparyan,rekalo}.
However, effects due to off-shellness of exchanged mesons have
been neglected in these studies. As in our previous studies of
$J/\psi$ absorption by nucleon \cite{liu1} and photoproduction of
$J/\psi$ on nucleons \cite{liu2}, we do not make the on-shell
approximation in evaluating the charmed meson production cross
section from proton-proton collisions.

This paper is organized as follows. In Section \ref{model}, we
introduce the interaction Lagrangians that are needed to evaluate
the cross sections for charm production from proton-proton
collisions. The two reactions $pp\to \bar{D}^0 p\Lambda_c$ and
$pp\to \bar{D}^{*0}p\Lambda_c$ are then
discussed in Section \ref{charm}. We show in this section the
amplitudes for these reactions and calculate their cross
sections due to contributions from pion, rho meson, $D$, and $D^*$
exchanges. The total cross section for charm production in
proton-proton collisions is given in Section \ref{total} and
compared to available experimental data. Finally, a brief summary
is given in Section \ref{summary}.

\section{the hadronic model}\label{model}

Possible reactions for charmed hadron production in
proton-proton collisions near threshold are $pp\to \bar
{D}^0p\Lambda_c^+$ and $pp\to \bar{D}^{*0}p\Lambda_c^+$. Cross
sections for these reactions can be evaluated using the same
Lagrangian introduced in Refs. \cite{liu1,liu2,lin2,di} for
studying charmed meson scattering by hadrons. This Lagrangian is
based on the gauged SU(4) flavor symmetry but with empirical
masses. The coupling constants are taken, if possible, from
empirical information. Otherwise, the SU(4) relations are used to
relate unknown coupling constants to known ones. Form factors are
introduced at interaction vertices with empirically determined
cutoff parameters.

\subsection{interaction Lagrangians}

From the formalism described in Refs. \cite{liu1,liu2,lin2,di}, the
interaction Lagrangian densities that are relevant to present study
are given as follows:
\begin{eqnarray}
{\cal L}_{\pi NN} & = & -ig_{\pi NN}\bar{N}\gamma_{5}
\vec{\tau}N\cdot\vec{\pi},\nonumber\\
{\cal L}_{\rho NN} & = & g_{\rho NN}\bar{N}\left(\gamma^{\mu}\vec{\tau}\cdot
\vec{\rho}_{\mu}+\frac{\kappa_{\rho}}{2m_{N}}\sigma_{\mu\nu}\vec{\tau}
\cdot\partial_{\mu}\vec{\rho}_{\nu}\right)N, \nonumber\\
{\cal L}_{\pi DD^{*}} & = & ig_{\pi DD^{*}}D^{*\mu}\vec{\tau}\cdot(\bar{D}
\partial_{\mu}\vec{\pi}-\partial_{\mu}\bar{D}\vec{\pi})+{\rm H.c.},
\nonumber\\
{\cal L}_{\rho DD} & = & ig_{\rho DD}(D\vec{\tau}\partial_{\mu}\bar{D}-
\partial_{\mu}D\vec{\tau}\bar{D})\cdot\vec{\rho}^{\mu},\nonumber\\
{\cal L}_{\rho D^{*}D^{*}} & = & ig_{\rho D^{*}D^{*}}[(\partial_{\mu}D^{*\nu}
\vec{\tau}\bar{D}^{*}_{\nu}-D^{*\nu}\vec{\tau}\partial_{\mu}
\bar{D}^{*}_{\nu})\cdot\vec{\rho}^{\mu} \nonumber\\
&+&(D^{*\nu}\vec{\tau}\cdot\partial_{\mu}\vec{\rho}^{\nu}
-\partial_{\mu}D^{*\nu}\vec{\tau}\cdot\vec{\rho}_{\nu})\bar{D}^{*\mu}
\nonumber\\
&+&D^{*\mu}(\vec{\tau}\cdot\vec{\rho}^{\nu}\partial_{\mu}
\bar{D}^{*}_{\nu}-\vec{\tau}\cdot\partial_{\mu}\vec{\rho}^{\nu}
\bar{D}^{*}_{\nu})],\nonumber\\
{\cal L}_{DN\Lambda_{c}} & = &ig_{DN\Lambda_{c}}
(\bar{N}\gamma_{5}\Lambda_{c}\bar{D}+D
\bar{\Lambda}_{c}\gamma_{5}N),\nonumber\\
{\cal L}_{D^{*}N\Lambda_{c}} & = &g_{D^{*}N\Lambda_{c}}
(\bar{N}\gamma_{\mu}\Lambda_{c}D^{*\mu}+\bar{D}^{*\mu}\bar{\Lambda}_{c}
\gamma_{\mu}N),\nonumber\\
{\cal
L}_{\pi\Lambda_c\Sigma_c}&=&ig_{\pi\Lambda_c\Sigma_c}\bar{\Lambda}_c
\gamma^5\vec{\Sigma}_c\cdot\vec{\pi}+{\rm H.c.}~,\nonumber\\
{\cal
L}_{\rho\Lambda_c\Sigma_c}&=&g_{\rho\Lambda_c\Sigma_c}\bar{\Lambda}_c
\gamma^\mu \vec{\Sigma}_c\cdot\vec{\rho_\mu}+{\rm H.c.}~,\nonumber\\
{\cal
L}_{DN\Sigma_c}&=&ig_{DN\Sigma_c}(\bar{N}\gamma^5\vec{\tau}\cdot
\vec{\Sigma}_c\bar{D}+D\vec{\tau}\cdot\bar{\vec{\Sigma}}_c\gamma^5
N),
\nonumber\\
{\cal L}_{D^* N\Sigma_c}&=&g_{D^* N\Sigma_c}(\bar{N}\gamma^\mu\vec{\tau}\cdot
\vec{\Sigma}_c\bar{D}^*_\mu+D^*_\mu\vec{\tau}\cdot\bar{\vec{\Sigma}}_c
\gamma^\mu N).
\end{eqnarray}
In the above, ${\vec\tau}$ are Pauli spin matrices, and
$\vec{\pi}$ and $\vec{\rho}$ denote, respectively, the pion and
rho meson isospin triplet, while $D=(D^+,D^0)$ and
$D^*=(D^{*+},D^{*0})$ denote, respectively, the pseudoscalar and
vector charmed meson doublets.

\subsection{coupling constants}

For coupling constants, we use the following empirical values:
$g_{\pi NN}=13.5$ \cite{pnn}, $g_{\rho NN}=3.25$, and $\kappa_\rho=6.1$
\cite{rnn}, and $g_{\pi DD^*}=5.56$ \cite{pdd}, and 
$g_{\rho DD}=g_{\rho D^*D^*}=2.52$
\cite{di}. Other coupling constants, which are not known empirically, are
obtained using SU(4) relations \cite{lin1,liu1,liu2,di}, i.e.,
\begin{eqnarray}
g_{D^{*}N\Lambda_c}&=&-\sqrt{3}g_{\rho NN}=-5.6,\nonumber\\
g_{DN\Lambda_c}&=&\frac{3-2\alpha_D}{\sqrt{3}}
g_{\pi NN}\simeq g_{\pi NN}=13.5,\nonumber\\
g_{\pi\Lambda_c\Sigma_c}&\simeq&-\frac{2\alpha_D}{\sqrt{3}}g_{\pi
NN},~~
g_{DN\Sigma_c}\simeq(1-2\alpha_D)g_{DN\Lambda_c},\nonumber\\
g_{D^*N\Sigma_c}&=&-g_{\rho NN}.
\end{eqnarray}
where $\alpha_D=D/(D+F)\simeq 0.64$ \cite{adelseck} with $D$ and $F$
being the coefficients for usual $D$-type and $F$-type couplings.

\subsection{form factors}

To take into account finite sizes of hadrons, form factors are
introduced at interaction vertices. In previous studies on $J/\psi$
absorption and charmed hadron scattering by hadrons, monopole form factors
have been used. Following the work on $J/\psi$ absorption by nucleons
\cite{liu1}, the form factors at $\pi NN$ and $\rho NN$ vertices are taken
to have the form:
\begin{eqnarray}\label{form1}
F_1(t)=\frac{\Lambda^2-m^2}{\Lambda^2-t},
\end{eqnarray}
with $t$ being the squared four momentum of exchanged pion or rho
meson, while those at $\pi DD^*$, $\rho DD$, $\rho D^*D^*$, 
$DN\Lambda_c$, $D^*N\Lambda_c$, $DN\Sigma_c$,
and $D^*N\Sigma_c$ vertices, that involve heavy virtual charm
mesons or baryons, are
\begin{eqnarray}
F_2({\bf q}^2)=\frac{\Lambda^2}{\Lambda^2+{\bf q}^2}.
\end{eqnarray}
with ${\bf q}$ being the three momentum transfer in the
center-of-mass frame for $t$ and $u$ channels or momentum of 
initial or final particles in center-of-mass frame for $s$
channel\cite{di}. As in Refs.\cite{pnn,rnn}, we take $\Lambda_{\pi
NN}=1.3$ GeV and $\Lambda_{\rho NN}=1.4$ GeV. For the cutoff parameters in
$F_2({\bf q}^2)$, they are taken to be 0.42 GeV. As discussed later
in Section \ref{charm}, this cutoff parameter is needed in a similar
hadronic model to reproduce the empirical cross section for
$pp\to K^+p\Lambda$ reaction at center-of-mass energy from threshold
to a few GeV,

\section{charmed hadron production from proton-proton collisions}
\label{charm}

In proton-proton collisions at low energies, charm production is
dominated by three particle final states. Two possible reactions
are $pp\to \bar{D}^0p\Lambda_c$ and $pp\to \bar{D}^{*0}p
\Lambda_c$. In the following, we discuss their contributions
separately.

\subsection{$pp\to \bar{D}^0 p\Lambda_c^+$}\label{dlambdac}

\begin{figure}[ht]
\centerline{\epsfig{file=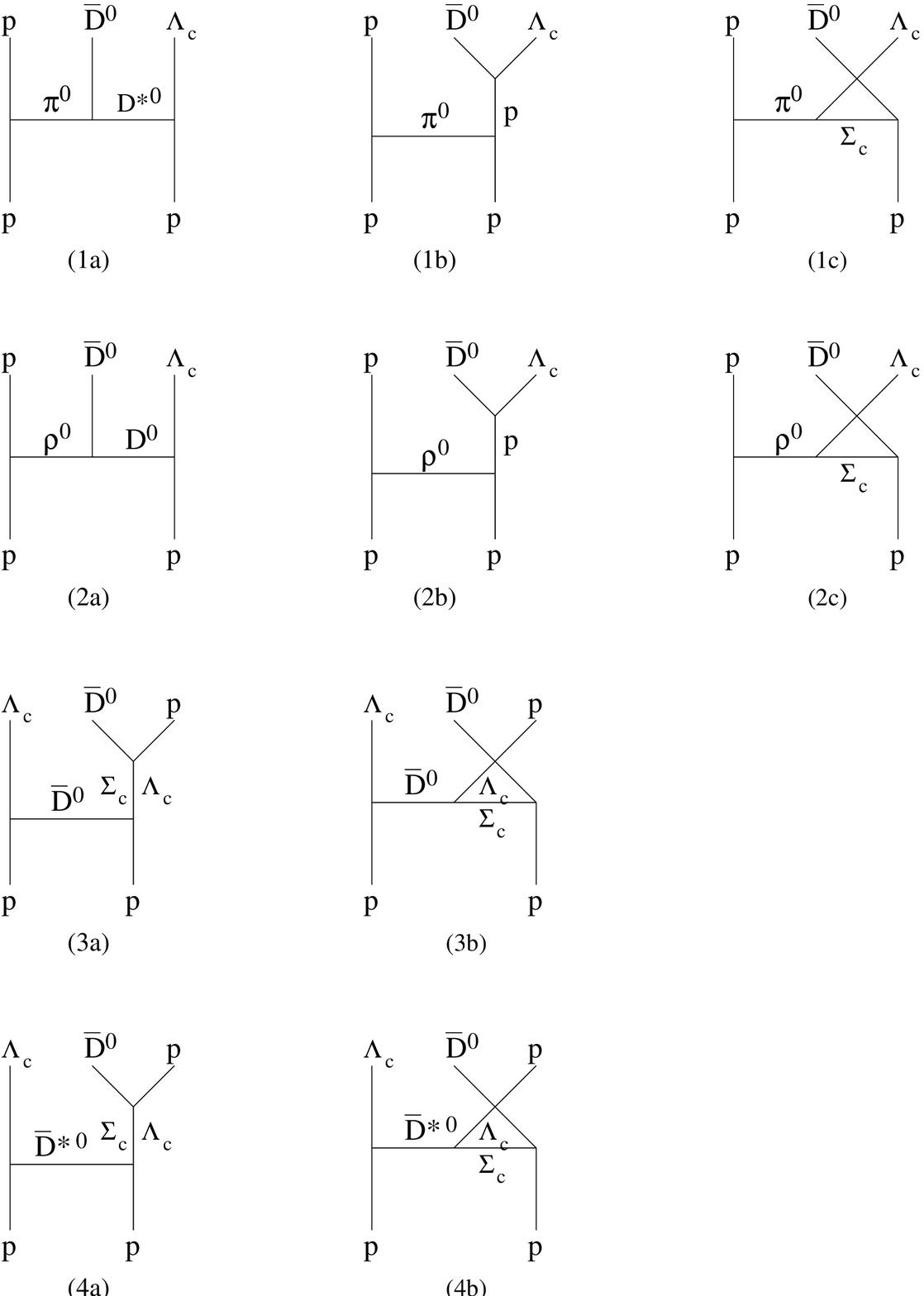,width=2.9in,height=3.9in,angle=0}}
\vspace{0.5cm}
\caption{Charmed hadron production from $pp\to \bar D^0p\Lambda_c^+$.} 
\label{diagramd}
\end{figure}

Diagrams for the reaction $pp\to \bar D^0p\Lambda_c^+$ are shown in
Fig. \ref{diagramd}. They involve the exchange of pion
($(1a)-(1c)$), rho meson ($(2a)-(2c)$), $D$ ($(3a)-(3b)$), and
$D^*$ ($(4a)-(4b)$). Amplitudes for the four processes are
given by
\begin{eqnarray}
{\cal M}_1 & = & -ig_{\pi NN}\bar{p}(p_{3})\gamma_{5}p(p_{1})
\frac{1}{t-m^{2}_{\pi}}\nonumber\\
&\times&({\cal M}_{1a}+{\cal M}_{1b}+{\cal M}_{1c}),\nonumber\\
{\cal M}_2 & = & g_{\rho NN}\bar{p}(p_{3})\left[\gamma^{\mu}
+i\frac{\kappa_{\rho}}{2m_{N}}\sigma^{\alpha\mu}
(p_{1}-p_{3})_{\alpha}\right]p(p_1)\nonumber\\
&\times& \left[-g_{\mu\nu}+\frac{(p_1-p_3)_{\mu}
(p_1-p_3)_{\nu}}{m^{2}_{\rho}}\right]\nonumber\\
&\times&\frac{1}{t-m^{2}_{\rho}}
({\cal M}^{\nu}_{2a}+{\cal M}^{\nu}_{2b}+{\cal M}^{\nu}_{2c}),\nonumber\\
{\cal M}_3 & = &
ig_{DN\Lambda_c}\bar{\Lambda}_c(p_{3})\gamma_{5}p(p_{1})
\frac{1}{t-m^{2}_{D}}\nonumber\\
&\times&({\cal M}_{3a}+{\cal M}_{3b}),\nonumber\\
{\cal M}_4 & = & g_{D^* N\Lambda_c}\bar{\Lambda}_c(p_{3})\gamma^{\mu}
p(p_{1})\nonumber\\
&\times&\left[-g_{\mu\nu}+\frac{(p_1-p_3)_{\mu}
(p_1-p_3)_{\nu}}{m^{2}_{D^*}}\right]\nonumber\\
&\times&\frac{1}{t-m^{2}_{D^*}} ({\cal M}^{\nu}_{4a}+{\cal
M}^{\nu}_{4b}),
\end{eqnarray}
where $p_1$ and $p_3$ are, respectively, four momenta of initial and final
baryons on the left side of a diagram, and $t=(p_1-p_3)^2$ is the square of
nucleon momentum transfer. The amplitudes $M_{ia}$, $M_{ib}$, and
$M_{ic}$ are for the subprocesses $\pi^0 p\to\bar
D^0\Lambda_c^+$, $\rho^0 p\to\bar D^0\Lambda_c^+$, $\bar D^0 p\to\bar
D^0p^+$, and $\bar D^{*0}p\to\bar D^0p$ involving exchanged virtual
mesons, and they are given explicitly by
\begin{eqnarray}
{\cal M}_{1a}&=& -g_{\pi
DD^*}g_{D^*N\Lambda_c}\frac{1}{q^2-m^2_{D^*}}
(k_1 +k_3)_\mu\nonumber\\
&\times&\left[g^{\mu\nu}-\frac{(k_1-k_3)^\mu(k_1-k_3)^\nu}{m^2_{D^*}}
\right]\bar{\Lambda}_c\gamma_\nu p,\nonumber\\
{\cal M}_{1b}&=&g_{\pi NN}g_{DN\Lambda_c}\frac{1}{s_1-m^2_N}
\bar{\Lambda}_c(m_N-{k\mkern-10mu/}_1-{k\mkern-10mu/}_2)p,\nonumber\\
{\cal M}_{1c}&=&g_{\pi\Lambda_c\Sigma_c}g_{DN\Sigma_c}
\frac{1}{u-m^2_{\Sigma_c}}
\bar{\Lambda}_c({k\mkern-10mu/}_2-{k\mkern-10mu/}_3-m_{\Sigma_c})p,\nonumber\\
{\cal M}^\mu_{2a}&=&ig_{DN\Lambda_c}g_{\rho DD}\frac{1}{q^2-m^2_D}
(2k_3-k_1)^\mu\bar{\Lambda}_c\gamma^5 p,\nonumber\\
{\cal M}^\mu_{2b}&=&ig_{\rho NN}g_{DN\Lambda_c}\frac{1}{s_1-m^2_N}
\bar{\Lambda}_c\gamma^5({k\mkern-10mu/}_1+{k\mkern-10mu/}_2+m_N)\nonumber\\
&\times&\left(\gamma^\mu+i\frac{\kappa_\rho}{2m_N}\sigma^{\beta\mu}k_{1\beta}
\right)p,\nonumber\\
{\cal M}^\mu_{2c}&=&ig_{\rho\Lambda_c\Sigma_c}g_{DN\Sigma_c}
\frac{1}{u-m^2_{\Sigma_c}}\nonumber\\
&\times&\bar{\Lambda}_c\gamma^\mu
({k\mkern-10mu/}_2-{k\mkern-10mu/}_3+m_{\Sigma_c})\gamma^5 p,\nonumber\\
{\cal M}_{3a}&=&g^2_{DN\Lambda_c}\frac{1}{s_1-m^2_{\Lambda_c}}
\bar{p}({k\mkern-10mu/}_1+{k\mkern-10mu/}_2-m_{\Lambda_c})p,\nonumber\\
{\cal M}_{3b}&=&g^2_{DN\Lambda_c}\frac{1}{u-m^2_{\Lambda_c}}
\bar{p}({k\mkern-10mu/}_2-{k\mkern-10mu/}_3-m_{\Lambda_c})p,\nonumber\\
{\cal M}^\mu_{4a}&=&ig_{D^* N\Lambda_c}g_{DN\Lambda_c}
\frac{1}{s_1-m^2_{\Lambda_c}}\nonumber\\
&\times&\bar{p}\gamma^5
({k\mkern-10mu/}_1+{k\mkern-10mu/}_2+m_{\Lambda_c})\gamma^\mu p,\nonumber\\
{\cal M}^\mu_{4b}&=&ig_{D^* N\Lambda_c}g_{DN\Lambda_c}
\frac{1}{u-m^2_{\Lambda_c}}\nonumber\\
&\times&\bar{p}\gamma^\mu
({k\mkern-10mu/}_2-{k\mkern-10mu/}_3+m_{\Lambda_c})\gamma^5 p.
\end{eqnarray}
Here, $k_1$ and $k_3$ are momenta of initial and final mesons, while
$k_2$ and $k_4$ are momenta of initial and final baryons in the
two-body subprocesses; and $q^2=(k_1-k_3)^2$ is the square of meson 
momentum transfer.

There is no interference between amplitudes involving 
exchange of pseudoscalar and vector mesons. Interferences
between amplitudes involving exchange of pion and $D$
meson as well as those between rho meson and $D^*$ are unimportant
due to the large mass difference between light and heavy mesons.
Neglecting these interferences, the total cross section for the
reaction $pp\to \bar D^0p\Lambda_c^+$ is then given by the sum of
the cross sections for the four processes in Fig. \ref{diagramd}
and can be expressed in terms of off-shell cross sections for
the subprocesses $\pi^0 p\to \bar D^0\Lambda_c^+$, $\rho^0 p\to \bar
D^0\Lambda_c^+$, $\bar D^0 p\to \bar D^0p$, and $\bar D^{*0}p\to
\bar D^0p$. Following the method of Ref. \cite{liu2} for the
reaction $J/\psi N\to D(D^*)\bar D(\bar D^*)N$, the spin-averaged
differential cross section for the reaction $pp\to \bar
D^0p\Lambda_c^+$ can be written as
\begin{eqnarray}
&&\frac{d\sigma_{pp\to \bar{D}^0p\Lambda_c^+}}{dtds_1}\nonumber\\
&& = \frac{g^{2}_{\pi NN}}
{16\pi^{2}sp^{2}_{i}}k\sqrt{s_{1}}(-t)\frac{1}
{(t-m^{2}_{\pi})^{2}}\sigma_{\pi^0 p\to \bar{D}^0\Lambda_c^+}(s_{1},t),
\nonumber\\
& &+ \frac{3g^{2}_{\rho NN}}{32\pi^{2}sp^{2}_{i}}k\sqrt{s_{1}}
\frac{1}{(t-m^{2}_{\rho})^{2}}
\left[4(1+\kappa_{\rho})^2\right .\nonumber\\
&&\times(-t-2m^{2}_{N})\kappa^{2}_{\rho}\frac{(4m^{2}_{N}-t)^{2}}{2m^{2}_{N}}
+4(1+\kappa_{\rho})\nonumber\\
&&\times\left .\kappa_{\rho}(4m^{2}_{N}-t)\right]
\sigma_{\rho^0 p\to \bar{D}^0\Lambda_c^+}(s_{1},t),\nonumber\\
&&+\frac{g^{2}_{DN\Lambda_c}}{16\pi^{2}sp^{2}_{i}}k\sqrt{s_{1}}
[-t+(m_N-m_{\Lambda_c})^2]\frac{1}{(t-m^{2}_{D})^{2}}\nonumber\\
&&\times\sigma_{\bar{D}^0p\to \bar{D}^0p}(s_{1},t)\nonumber\\
&&+ \frac{3g^{2}_{D^* N\Lambda_c}}{32\pi^{2}sp^{2}_{i}}k\sqrt{s_{1}}
\frac{1}{(t-m^{2}_{D^*})^{2}}\left[-4t+4(m_{\Lambda_c}-m_N)^2\right.
\nonumber\\
&&-8m_{\Lambda_c}m_N+\frac{2(m^2_N -m^2_{\Lambda_c}-t)(m^2_N -m^2_{\Lambda_c}
+t)}{m^2_{D^*}}\nonumber\\
&&\left.+\frac{2((m_{\Lambda_c}-m_N)^2+t)t}{m^2_{D^*}}\right]
\sigma_{\bar{D}^{*0} p\to \bar{D}^0p}(s_{1},t).\label{pp2lcd}
\end{eqnarray}
In the above, $p_i$ is the center-of-mass momentum of two
initial protons, $t$ is the squared four momentum transfer of exchanged
meson, $s$ is the squared center-of-mass energy, and $s_{1}$ and
$k$ are, respectively, the squared invariant mass and
center-of-mass momentum of exchanged meson and the nucleon in the
subprocesses. We have also included a factor of two to take into
account contributions from interchanging two initial protons.

\begin{figure}[ht]
\centerline{\epsfig{file=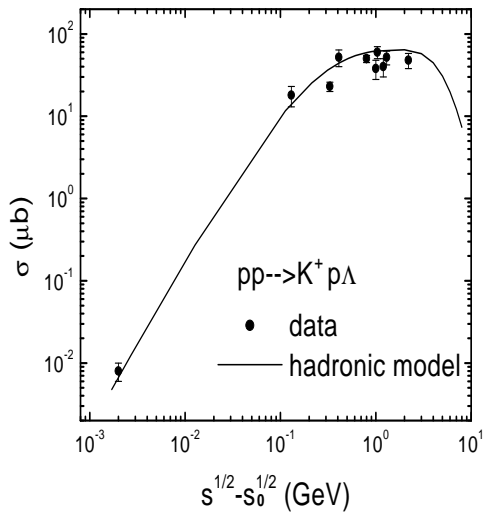,width=3in,height=3in,angle=-90}}
\caption{Cross section for kaon production from
the reaction $pp\to K^+p\Lambda$ with cutoff parameter $\Lambda=0.42$
GeV in the form factors at interaction vertices involving exchange of
strange mesons. Filled circles are experimental data taken
from Ref.\protect\cite{gqli}}
\label{pp}
\end{figure}

Since the charmed hadron production cross sections is sensitive to 
the value of cutoff paramters in the form factors at interaction 
vertices involving virtual charmed mesons and baryons, it is 
necessary to constraint this cutoff paramter empirically. Without
exclusive cross sections available for chamred hadron production
from proton-proton scattering, we resort to strange hadron production.
Usinig the same hadronic model for kaon production from the reaction
$pp\to K^+p\Lambda$, this reaction can be described by similar
diagrams in Fig.\ref{diagramd} for the reaction $pp\to \bar D^0p\Lambda_c^+$
with $D^0$ and $\Lambda_c$ replaced by $K^+$ and $\Lambda$,
respectively, in the final states. Also, the exchanged $\bar D^0$ in 
diagrams (3a) and (3b) as well as $\bar D^{0*}$ in diagrams (4a) and
(4b) are replaced by $K$ and $K^*$, respectively, while intermediate
off-shell charmed baryons are replaced by strange baryons.
With empirical coupling constants $g_{\pi KK^*}=3.25$
and $g_{\rho KK}=3.25$, as well as others determined via SU(3) relations 
\cite{changhui}, the measured cross section can be reproduced with a 
cutoff parameter $\Lambda=0.42$ GeV in the form factors 
$F_2({\bf q}^2)$ at vertices involving virtual strange mesons
and baryons, as shown in Fig.\ref{pp}.

\begin{figure}[ht]
\centerline{\epsfig{file=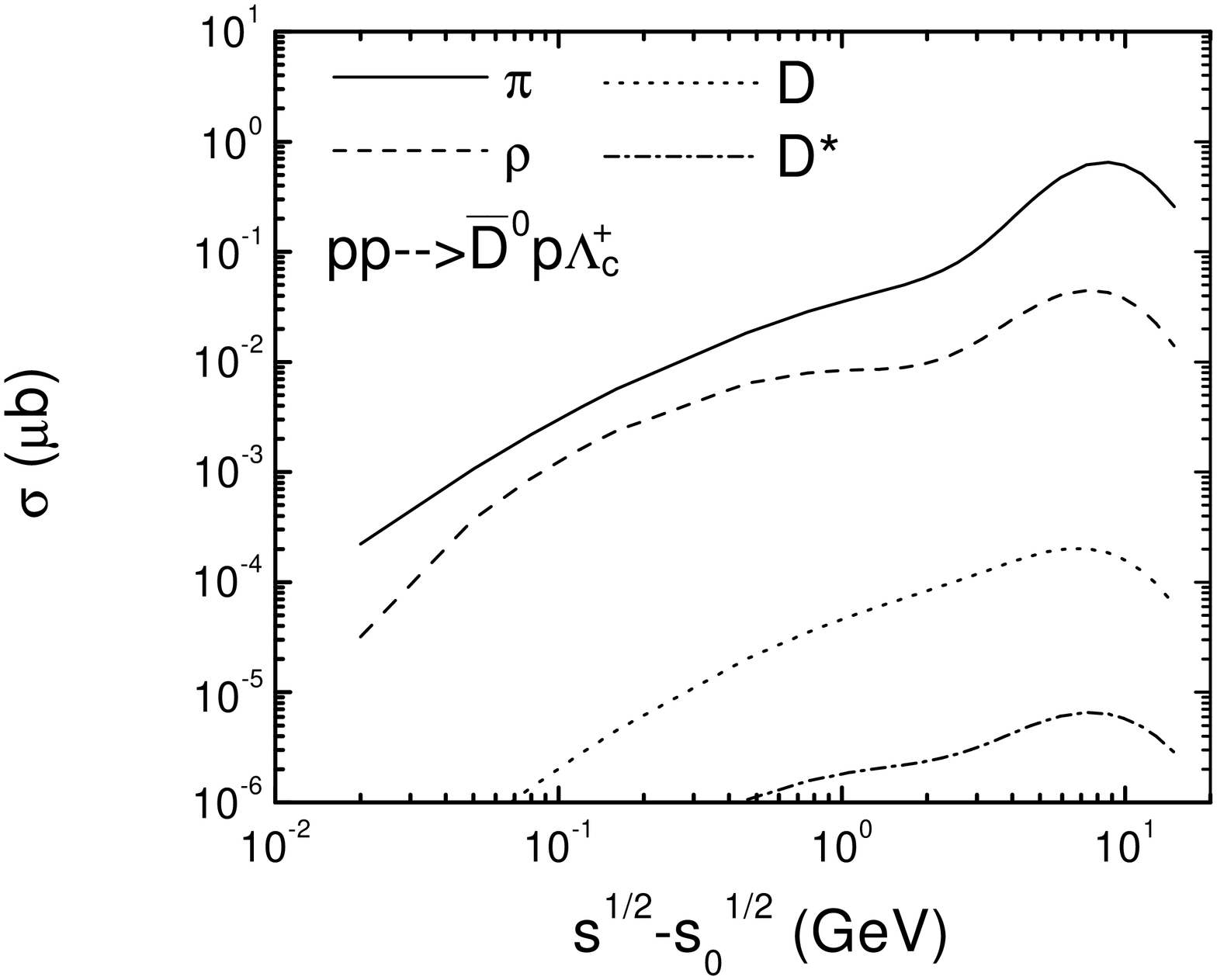,width=3in,height=3in,angle=-90}}
\caption{Cross sections for charmed hadron
production from the reaction $pp\to \bar D^0p\Lambda_c^+$ due to
pion (solid curve), rho meson (dashed curve), $D$ (dotted curve),
and $D^*$ (dash-dotted curve).} \label{crossd}
\end{figure}

Assuming that the same cutoff parameter $\Lambda=0.42$ GeV is applicable 
at vertices involving virtual charmed mesons and baryons in charmed
hadron production from proton-proton reactions, resulting cross
sections for the reaction $pp\to\bar D^0p\Lambda_c$ from the four 
possible processes of pion
(solid curve), rho (dashed curve), $D$ (dotted curve), and $D^*$
(dash-dotted curve) exchanges as functions of center-of-mass
energy are shown in Fig.\ref{crossd}. It is seen that contributions
from light meson exchange are more important than those from heavy
meson exchange.  Although we consider diagrams (1a) and (2a) in
Fig.\ref{diagramd} as exchange of pion and rho meson, respectively,
they actually involve exchange of heavy $D^*$ and $D$ mesons in the 
subprocess $\pi^0p\to\bar D^0\Lambda_c^+$ and 
$\rho^0p\to\bar D^0\Lambda_c^+$, respectively.
Our results that main contributions to the reaction
$pp\to p\bar D^0\Lambda_c^+$ are due to exchange of light mesons are not
inconsistent with conclusions in Ref.\cite{gasparyan} that this
reaction is dominated by heavy $D$ meson exchange.

\begin{figure}[ht]
\centerline{\epsfig{file=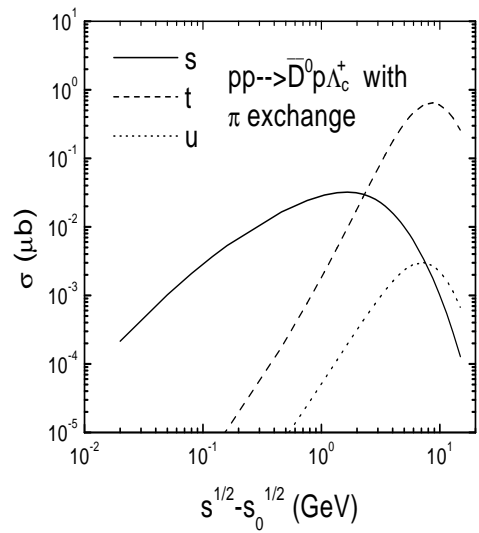,width=3in,height=3in,angle=-90}}
\caption{Partial cross sections for $pp\to\bar D^0p\Lambda_c^+$ due to
contributions from different channels.}
\label{compare}
\end{figure}

To see the relative contributions from $s$, $t$, and $u$ channel
diagrams in Fig.\ref{diagramd}, we show in Fig.\ref{compare}
the partial cross sections due to diagrams (1a), (1b), and (1c).
It is seen that the $t$ channel diagram (1a) dominates charmed hadron
production cross section at high energies while the $s$ channel
diagram (1b) is most important near threshold. The contribution from the
$u$ channel diagram (1c) is much smaller than those from other two diagrams.
Except near threshold, our results are thus similar to those found
in Ref. \cite{gasparyan}, which uses the on-shell approximation for
the subprocess $\pi p\to\bar D^0\Lambda_c^+$ and does not include $s$ and $u$
channel diagrams.

\subsection{$pp\to \bar D^{*0}p\Lambda_c^+$}

\begin{figure}[ht]
\centerline{\epsfig{file=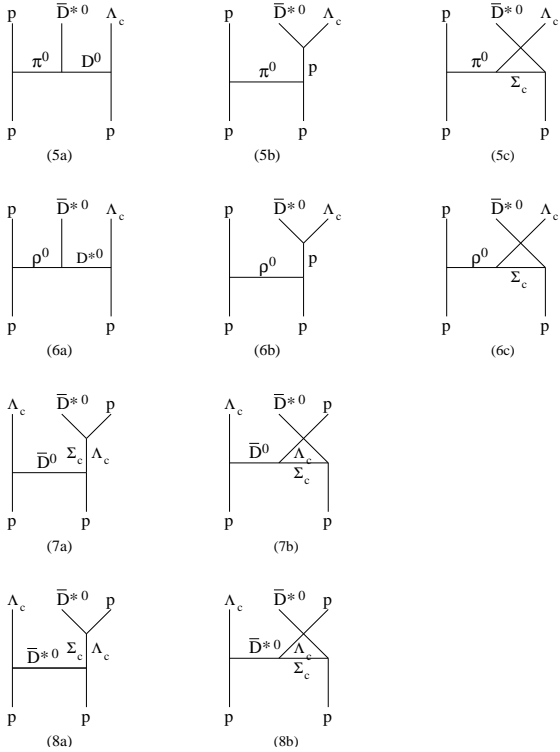,width=2.9in,height=3.9in,angle=0}}
\vspace{0.5cm}
\caption{Charmed hadron production from $pp\to \bar
D^{*0}p\Lambda_c^+$.} \label{diagramds}
\end{figure}

For charm production from proton-proton collisions with $\bar
D^{*0}p\Lambda_c^+$ in the final state, relevant diagrams are shown in
Fig. \ref{diagramds}. As for the reaction $pp\to \bar
D^{*0}p\Lambda_c^+$, this reaction can proceed through pion, rho meson,
$D$, and $D^*$ exchanges. Amplitudes for the four processes can be
evaluated with the interaction Lagrangians given in Section
\ref{model}, and they are given by
\begin{eqnarray}
{\cal M}_5& = & -ig_{\pi NN}\bar{p}(p_{3})\gamma_{5}p(p_{1})
\frac{1}{t-m^{2}_{\pi}}\nonumber\\
&\times&({\cal M}^\alpha_{5a}+{\cal M}^\alpha_{5b}+
{\cal M}^\alpha_{5c})\epsilon_\alpha\nonumber\\
{\cal M}_6 & = & g_{\rho NN}\bar{p}(p_{3})\left[\gamma^{\mu}
+i\frac{\kappa_{\rho}}{2m_{N}}\sigma^{\alpha\mu}
(p_{1}-p_{3})_{\alpha}\right]p(p_1)\nonumber\\
&\times& \left[-g_{\mu\nu}+\frac{(p_1-p_3)_{\mu}
(p_1-p_3)_{\nu}}{m^{2}_{\rho}}\right]\nonumber\\
&\times&\frac{1}{t-m^{2}_{\rho}} ({\cal M}^{\nu\alpha}_{6a}+{\cal
M}^{\nu\alpha}_{6b}
+{\cal M}^{\nu\alpha}_{6c})\epsilon_\alpha,\nonumber\\
{\cal M}_7 & = &
ig_{DN\Lambda_c}\bar{\Lambda}_c(p_{3})\gamma_{5}p(p_{1})
\frac{1}{t-m^{2}_{D}}\nonumber\\
&\times&({\cal M}^\alpha_{7a}+{\cal M}^\alpha_{7b})\epsilon_\alpha,\nonumber\\
{\cal M}_8 & = & g_{D^*
N\Lambda_c}\bar{\Lambda}_c(p_{3})\gamma^{\mu}
p(p_{1})\nonumber\\
&\times&\left[-g_{\mu\nu}+\frac{(p_1-p_3)_{\mu}(p_1-p_3)_{\nu}}{m^{2}_{D^*}}
\right]\nonumber\\
&\times&\frac{1}{t-m^{2}_{D^*}} ({\cal M}^{\nu\alpha}_{8a}+{\cal
M}^{\nu\alpha}_{8b})\epsilon_\alpha,
\end{eqnarray}
where $p_1$ and $p_3$ are again, respectively, four momenta of initial 
and final baryons on the left side of a diagram and $\epsilon_\alpha$ 
denotes the polarization vector of $D^*$ meson in final state.

Expressions for individual amplitudes can be written as follows:
\begin{eqnarray}
{\cal M}^\mu_{5a}&=&-ig_{\pi DD^*}g_{DN\Lambda_c}
\frac{1}{q^2-m^2_D}(2k_1-k_3)^\mu\bar{\Lambda}_c\gamma_5 p,\nonumber\\
{\cal M}^\mu_{5b}&=&-ig_{\pi NN}g_{D^*N\Lambda_c}\frac{1}{s_1-m^2_N}\nonumber\\
&\times&\bar{\Lambda}_c\gamma^\mu({k\mkern-10mu/}_1
+{k\mkern-10mu/}_2+m_N)\gamma^5 p,\nonumber\\
{\cal M}^\mu_{5c}&=&ig_{\pi\Lambda_c\Sigma_c}g_{D^* N\Sigma_c}
\frac{1}{u-m^2_{\Sigma_c}}\nonumber\\
&\times&\bar{\Lambda}_c\gamma^5({k\mkern-10mu/}_2
-{k\mkern-10mu/}_3+m_{\Sigma_c})\gamma^\mu p,\nonumber\\
{\cal M}^{\mu\nu}_{6a}&=&g_{D^* N\Lambda_c}g_{\rho D^* D^*}
\frac{1}{q^2-m^2_{D^*}}\nonumber\\
&\times&\left[g_{\alpha\beta}-\frac{(k_1-k_3)_\alpha
(k_1-k_3)_\beta}{m^2_{D^*}}\right]\bar{\Lambda}_c\gamma^\alpha p\nonumber\\
&\times&[2k^\nu_1 g^{\beta\mu}-(k_1+k_3)^\beta g^{\mu\nu}
+2k^\mu_3 g^{\beta\nu}],\nonumber\\
{\cal M}^{\mu\nu}_{6b}&=&g_{\rho
NN}g_{D^*N\Lambda_c}\frac{1}{s_1-m^2_N}
\bar{\Lambda}_c\gamma^\nu({k\mkern-10mu/}_1+{k\mkern-10mu/}_2+m_N)\nonumber\\
&\times&\left(\gamma^\mu+i\frac{\kappa_\rho}{2m_N}\sigma^{\beta\mu}k_{1\beta}
\right)p,\nonumber\\
{\cal M}^{\mu\nu}_{6c}&=&g_{\rho\Lambda_c\Sigma_c}g_{D^* N\Sigma_c}
\frac{1}{u-m^2_{\Sigma_c}}\nonumber\\
&\times&\bar{\Lambda}_c\gamma^\mu({k\mkern-10mu/}_2
-{k\mkern-10mu/}_3+m_{\Sigma_c})\gamma^\nu p.\nonumber\\
{\cal M}^\mu_{7a}&=&ig_{DN\Lambda_c}g_{D^*N\Lambda_c}
\frac{1}{s_1-m^2_{\Lambda_c}}\nonumber\\
&\times&\bar{p}\gamma^\mu({k\mkern-10mu/}_1
+{k\mkern-10mu/}_2+m_{\Lambda_c})\gamma^5 p,\nonumber\\
{\cal M}^\mu_{7b}&=&ig_{DN\Lambda_c}g_{D^* N\Lambda_c}
\frac{1}{u-m^2_{\Lambda_c}}\nonumber\\
&\times&\bar{p}\gamma^5({k\mkern-10mu/}_2
-{k\mkern-10mu/}_3+m_{\Lambda_c})\gamma^\mu p,\nonumber\\
{\cal M}^{\mu\nu}_{8a}&=&g^2_{D^*
N\Lambda_c}\frac{1}{s_1-m^2_{\Lambda_c}}\bar{p}
\gamma^\nu({k\mkern-10mu/}_1+{k\mkern-10mu/}_2+m_{\Lambda_c})\gamma^\mu
p,\nonumber\\
{\cal M}^{\mu\nu}_{8b}&=&g^2_{D^*
N\Lambda_c}\frac{1}{u-m^2_{\Lambda_c}}
\bar{p}\gamma^\mu({k\mkern-10mu/}_2-{k\mkern-10mu/}_3
+m_{\Lambda_c})\gamma^\nu p.
\end{eqnarray}

As in the case of charmed hadron production from the reaction
$pp\to \bar D^0p\Lambda_c^+$, total cross section for the
reaction $pp\to \bar D^{*0}p\Lambda_c^+$ can be expressed in terms
of off-shell cross sections for the subprocesses $\pi^0 p\to
\bar D^{*0}\Lambda_c^+$, $\rho^0 p\to \bar D^{*0}\Lambda_c^+$, $\bar
D^0 p\to \bar D^{*0}p$, and $\bar D^{*0}p\to \bar D^{*0}p$. In
this case, the spin averaged differential cross section is
\begin{eqnarray}
&&\frac{d\sigma_{pp\to \bar{D}^0p\Lambda_c}}{dtds_1}\nonumber\\
&&=\frac{g^{2}_{\pi
NN}}{16\pi^{2}sp^{2}_{i}}k\sqrt{s_{1}}(-t)\frac{1}
{(t-m^{2}_{\pi})^{2}}\sigma_{\pi^0p\to\bar{D}^{*0}\Lambda_c^+}(s_{1},t), 
\nonumber\\
& &+ \frac{3g^{2}_{\rho NN}}{32\pi^{2}sp^{2}_{i}}k\sqrt{s_{1}}
\frac{1}{(t-m^{2}_{\rho})^{2}}
\left[4(1+\kappa_{\rho})^2\right.\nonumber\\
&&\times(-t-2m^{2}_{N})\kappa^{2}_{\rho}\frac{(4m^{2}_{N}-t)^{2}}{2m^{2}_{N}}
+4(1+\kappa_{\rho})\nonumber\\
&&\times\left .\kappa_{\rho}(4m^{2}_{N}-t)\right]
\sigma_{\rho^0 p\to \bar{D}^{*0}\Lambda_c^+}(s_{1},t)\nonumber\\
&&+\frac{g^{2}_{DN\Lambda_c}}{16\pi^{2}sp^{2}_{i}}k\sqrt{s_{1}}
(-t+(m_N-m_{\Lambda_c})^2)\nonumber\\
&&\frac{1}{(t-m^{2}_{D})^{2}}\sigma_{\bar{D}^0p\to\bar{D}^{*0}p}(s_{1},t)
\nonumber\\
& &+ \frac{3g^{2}_{D^* N\Lambda_c}}{32\pi^{2}sp^{2}_{i}}k\sqrt{s_{1}}
\frac{1}{(t-m^{2}_{D^*})^{2}}\nonumber\\
&&\left[-4t+4(m_{\Lambda_c}-m_N)^2-8m_{\Lambda_c}m_N\right.\nonumber\\
&&+\frac{2(m^2_N-m^2_{\Lambda_c}-t)(m^2_N -m^2_{\Lambda_c}+t)}{m^2_{D^*}}
\nonumber\\
&&\left.+\frac{2((m_{\Lambda_c}-m_N)^2+t)t}{m^2_{D^*}}\right]\nonumber\\
&&\times\sigma_{\bar{D}^{*0} p\to \bar{D}^{*0}p}(s_{1},t).
\end{eqnarray}

\begin{figure}[ht]
\centerline{\epsfig{file=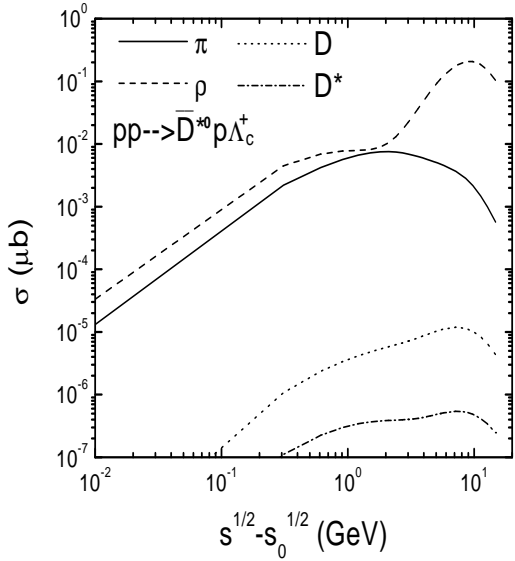,width=3.2in,height=3.2in,angle=-90}}
\caption{Cross sections for charmed hadron production from $pp\to
\bar D^{*0}p\Lambda_c^+$ due to pion (solid curve), rho meson
(dashed curve), $D$ (dotted curve), and $D^*$ (dash-dotted
curve).} \label{crossds}
\end{figure}

Using coupling constants and cutoff parameters introduced in
Section \ref{model}, we have evaluated the cross section for the
reaction $pp\to \bar D^{*0}p\Lambda_c^+$. In Fig. \ref{crossds}, we
show contributions from pion (solid curve), rho meson (dashed
curve), $D$ (dotted curve), and $D^*$ (dash-dotted curve)
exchanges as functions of center-of-mass energy. As for the
reaction $pp\to \bar D^0p\Lambda_c^+$, light meson exchanges are
more important than those from heavy meson exchanges. However, the
contribution from rho exchange is larger than that from pion
exchange, which is opposite to that in the reaction $pp\to\bar
D^0\Lambda_c^+$, as a result of couplings involving three vector
mesons, which are absent in the latter reaction.

\section{Total Cross Section for charmed hadron production in
proton-proton collisions}\label{total}

\begin{figure}[ht]
\centerline{\epsfig{file=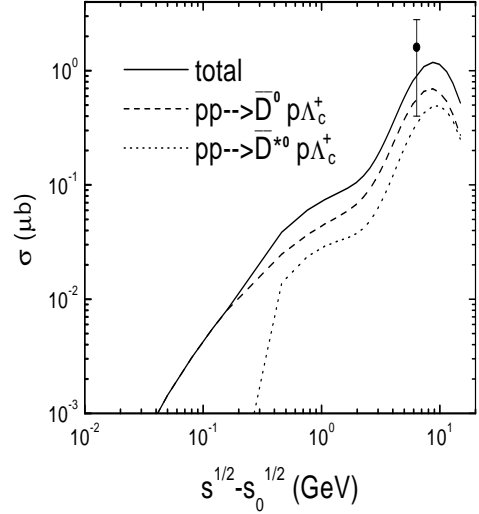,width=3in,height=3in,angle=-90}}
\caption{Cross sections for charmed hadron
production from proton-proton collisions. Dashed and dotted curves
are for $pp\to \bar D^0p\Lambda_c^+$ and $pp\to \bar
D^{*0}p\Lambda_c^+$, respectively, while the total cross section is
shown by solid curve. The threshold energy $s_0$ refers to that
of the reation $pp\to \bar D^0p\Lambda_c^+$. Experimental data are 
shown by filled circles \protect\cite{data}.} \label{crosst}
\end{figure}

The total cross section for charm production from proton-proton
collisions is shown in Fig.\ref{crosst} as a function of
center-of-mass energy (solid curve). It's value at center-of-mass energy 
of 11.5 GeV is about 1 $\mu$b and is within the uncertaity of
measured inclusive charm production cross section, which is about
2 $\mu$b as shown by solid circles with error bar \cite{data}.
The cross section decreases as energy drops and is about 1 nb 
at 40 MeV above threshold. Also shown in Fig.\ref{crosst} are the 
cross section for the reactions $pp\to \bar D^0p\Lambda_c^+$ 
(dashed curve) and $pp\to \bar D^{*0}p\Lambda_c^+$ (dotted curve), 
and it is seen that the former is somewhat larger than the latter.

\section{summary}\label{summary}

Using a hadronic model based on SU(4) flavor invariant Lagrangian
with empirical masses and coupling constants, we have studied charmed
hadron production from proton-proton collisions through the reactions
$pp\to\bar D^0p\Lambda_c^+$ and $pp\to\bar D^{*0}p\Lambda_c^+$. These reactions
involve exchange of pion, rho meson, $D$, and $D^*$, and their cross
sections can be expressed in terms of the cross sections for the off-shell
processes $Mp\to\bar D^0\Lambda_c^+$ and $Mp\to\bar
D^{*0}\Lambda_c^+$, where $M$ denotes one of the above exchanged
off-shell mesons. With cutoff parameters of form factors
adjusted to fit the cross section for strange hadron production in
proton-proton reactions, the resulting cross section for charmed hadron
production from proton-proton collisions at center-of-mass energy
of 11.5 GeV is consistent with available experimental data.
The predicted cross section at 40 MeV above threshold is about 
1 nb. Our results will be useful for the experiments to be
carried out at proposed accelerator at the German Heavy Ion Research 
Center \cite{gsi}.

\section*{acknowledgment}

This paper is based on work supported by the National Science Foundation
under Grant No. PHY-0098805 and the Welch Foundation under Grant
No. A-1358. SHL is also supported in part by the KOSEF under Grant
No. 1999-2-111-005-5 and by the Korea Research Foundation under Grant
No. KRF-2002-015-CP0074.

\end{multicols}

\end{document}